\documentclass{emulateapj}

\usepackage{natbib}
\usepackage{apjfonts}

\usepackage{amsmath}
\usepackage{rotating}

\newcommand{\gx}{GX~339$-$4}

\shorttitle{Going with the flow: corona-jet unification?}
\shortauthors{Markoff, Nowak \& Wilms}

\begin{document}

\title{Going with the flow: can the base of jets subsume the role of
compact accretion disk coronae?} 

\author{Sera Markoff\altaffilmark{1} and Michael A. Nowak}
\affil{Massachusetts Institute of Technology, Kavli Institute for
  Astrophysics and Space Research, Rm. NE80-6035, Cambridge, MA 02139}
\email{sera,mnowak@space.mit.edu}

\and

\author{J\"orn Wilms}
\affil{Department of Physics, University of Warwick, Coventry, CV4
  7AL, United Kingdom}\email{j.wilms@warwick.ac.uk}

\altaffiltext{1}{NSF Astronomy \& Astrophysics Postdoctoral Fellow}

\begin{abstract}

The hard state of X-ray binaries (XRBs) is characterized by a power
law spectrum in the X-ray band, and a flat/inverted radio/IR spectrum
associated with occasionally imaged compact jets.  It has generally
been thought that the hard X-rays result from Compton upscattering of
thermal accretion disk photons by a hot, coronal plasma whose
properties are inferred via spectral fitting.  Interestingly, these
properties---especially those from certain magnetized corona
models---are very similar to the derived plasma conditions at the jet
footpoints.  Here we explore the question of whether the `corona' and
`jet base' are in fact related, starting by testing the strongest
premise that they are synonymous.  In such models, the radio through
the soft X-rays are dominated by synchrotron emission, while the hard
X-rays are dominated by inverse Compton at the jet base -- with both
disk and synchrotron photons acting as seed photons.  The conditions
at the jet base fix the conditions along the rest of the jet, thus
creating a direct link between the X-ray and radio emission.  We also
add to this model a simple iron line and convolve the spectrum with
neutral reflection.  After forward-folding the predicted spectra
through the detector response functions, we compare the results to
simultaneous radio/X-ray data obtained from the hard states of the
Galactic XRBs \gx\ and Cygnus X-1.  Results from simple Compton corona
model fits are also presented for comparison.  We demonstrate that the
jet model fits are statistically as good as the single-component
corona model X-ray fits, yet are also able to address the simultaneous
radio data.

\end{abstract}


\section{Introduction}

Bipolar plasma outflows, loosely termed jets, are a feature common to
a variety of different astrophysical objects.  Jets are observed in
accreting compact objects of all scales, as well as at stellar birth
and death.  They seem equally able to form out of massive accretion
disks as well as from quasi-spherically collapsing plasma, which
suggests that their creation is a basic byproduct of some routinely
occurring ingredients.  These necessary inputs seem to be rotation,
collapsing or infalling/accreting plasma, and magnetic fields.  Jets
likely are formed at least in part from a necessity to shed the system
of excess angular momentum \citep[e.g.,][]{MeierNakamura2004}.  The
exact details of their formation, as well as the nature of the
coupling between the infalling/collapsing plasma and the outflow is
still an active area of controversy and research on many fronts.
However, the fact that so many types of systems share this mechanism
extends hope that clues gained from one class of objects can be
applied to the others in the search for a unified picture.

The advantage of looking for answers in X-ray binary (XRB) jets is
twofold.  First, their formation seems to be a recurrent phenomenon on
time scales we can observe repeatedly over the course of our own human
lifetimes.  For instance in the Galactic XRB GX~339$-$4 and Cyg~X-1,
accretion without observable jets proceeds into accretion with
discernible jets and back again every few years
\citep{Fender2003,Pottschmidtetal2003,Gleissneretal2004,Homanetal2005}.
Thus the same object can provide tests of theories developed based on
its earlier activity.  Secondly, the existence of quality simultaneous
broadband data for many sources linking the ``traditional'' jet
outflow bands (radio and likely also infrared, or IR) and inflow bands
(X-ray and optical) gives us a way of studying the relationship and
interplay between the two sides of accretion.  We use the word
``traditional'' to refer to the picture which existed up until
recently that the jets and accretion flow radiate in different energy
bands and can generally be studied as distinct phenomena.  This
picture has rapidly evolved in the last few years, beginning with the
discovery of an unexpectedly intimate connection between the low and
high frequency wavebands in the hard state (HS) of XRBs (see, e.g.,
\citealt{McClintockRemillard2003}, for a detailed definition).

Using radio observations in conjunction with 1.3-12.2\,keV data from
the All Sky Monitor (ASM) on-board the Rossi X-ray Timing Explorer
\citep[RXTE;][]{Levineetal1996}, \cite{Hannikainenetal1998} first
discovered a near-linear correlation between the radio and X-ray
emission of GX~339$-$4.  These studies were followed by joint radio
observations and pointed X-ray observations with RXTE
\citep{Wilmsetal1999,NowakWilmsDove2002} .  This intensive
simultaneous monitoring showed that, in fact, the correlation is
non-linear, with $L_{\rm R} \propto L_{\rm X}^{0.7}$
\citep{Corbeletal2000,Corbeletal2003}.  Additionally, this non-linear
correlation holds over orders of magnitude changes in the source
luminosity with time.  In fact, even after fading into quiescence,
GX~339$-$4 seemingly returns to the same correlation in subsequent
outbursts \citep{Corbeletal2003}, although it is not clear whether the
correlation always maintains the same normalization in each instance
\citep{Nowaketal2005}.

Despite some possible variation in its normalization, the correlation
between radio and X-ray fluxes appears to be universal to all HS XRBs
with comparable broadband data \citep{GalloFenderPooley2003}.  The
$0.7$ slope of the $\log(L_{\rm R})$-$\log(L_{\rm X})$ correlation
follows directly from analytic predictions of scaling synchrotron jet
models \citep{FalckeBiermann1995,Markoffetal2003}, and can be
generalized for other X-ray emission processes in terms of their
dependence on the accretion rate \citep{HeinzSunyaev2003}.  The
correlation's normalization in the $(L_{\rm X},L_{\rm R})$ plane also
depends on the central engine mass.  When this scaling is accounted
for, unbeamed, extra-galactic, supermassive black holes sources agree
remarkably well with the same radio/X-ray correlation found in
Galactic, stellar-mass sources
\citep{MerloniHeinzDiMatteo2003,FalckeKoerdingMarkoff2004}.  The
underlying physics governing the connection between the radio/IR and
X-ray bands must therefore be fundamental to accreting sources
regardless of mass.

In \cite{MarkoffFalckeFender2001} we showed that jet synchrotron emission
could account for the broad continuum features of the simultaneous
radio through X-ray hard state observation of the Galactic XRB,
XTE~J1118+480.  In a later work, we showed that this same model could
also explain the broad spectral features of thirteen simultaneous or
quasi-simultaneous radio/X-ray observations of GX~339$-$4, and that
the radio/X-ray correlation was naturally produced by only changing
the power input into the jet \citep{Markoffetal2003}.  These models,
however, did not attempt to address the fine features present in the
X-ray spectrum which are the hallmarks of an optically thick accretion
disk: in particular fluorescent iron lines and a characteristic
flattening above 10\,keV attributed to Compton reflection
\citep{LightmanWhite1988,RossFabian1993}.  The reflection is generally
assumed to result from a hard X-ray continuum originating above the
cooler accretion disk.

We began to explore the interaction of the jet emission with the
putative accretion disk in \cite{MarkoffNowak2004}.  In the hard state
(HS), the spectrum is hard enough and the observed reflection
signature weak enough to be problematic for coronal models where there
is significant coverage of the disk by hard X-ray-emitting material
\citep[e.g.,][and references therein]{dove:97b}.  Various mechanisms
have been proposed to decrease both the cooling of the hard
X-ray-emitting material by soft disk photons and the fraction of
reflected X-rays.  These mechanisms include patchy coronae
\citep{stern:95b}, high disk ionization
\citep{ross:99a,Nayakshin2000,ballantyne:01a} and beaming of the
coronae away from the disk with mildly relativistic velocities
\citep{Beloborodov1999,MalzacBeloborodovPoutanen2001}.  This latter
approach seems extremely close in principle to the characteristics of
the base of a jet.

Compact accretion disk coronae are theoretical concepts based on
observations that suggest the presence of hot electrons radiating near
the inner parts of a thin accretion disk.  The existence of hot
electrons in the same region is also empirically required by the jets
we image in the radio wavebands.  The high brightness temperatures and
occasionally measured linear polarization \citep[e.g.,][]{Fender2001a}
argue strongly that nonthermal synchrotron is the dominant
radio-emitting process.  Because the base of the jet is significantly
more compact than the outer region which dominates the cm-bands,
conservation arguments imply an even hotter, denser medium near the
accretion disk.  We therefore wish to determine whether the base of
the jets can ``subsume'' the role of the corona.  This idea was 
proposed already back in \cite{Fenderetal1999b}, and by subsume we
mean provide the spectral characteristics (both direct and reflected)
which are traditionally attributed to a compact Comptonizing corona.
Thus we will compare coronal fits using a standard Comptonizing model
(see below) to fits made by our jet model.  Regardless of the ultimate
relationship determined to exist between the base of the jets and the
corona, this is a critical step in the road to understanding.  The
results of this study will provide valuable clues as to the nature
of the corona, both in terms of geometrical as well as internal
characteristics which differ from the base of the jets.

As an additional note, the concept of cyclo-synchrotron photons
feeding inverse Compton processes in a corona has existed for over 20
years \citep{Fabianetal1982}, inspired by multiwavelength observations
showing similar optical/X-ray behavior \citep{Motchetal1982}.  It has
recently been reconsidered for several sources
\citep{diMatteoCelottiFabian1997,MerlonidiMatteoFabian2000,
WardzinskiZdziarski2000} in the context of magnetic flares in the
corona.  While a valuable step towards understanding the role of
magnetic fields, these models do not address the radio emission from
the jets with the same model.

The remainder of this paper is structured as follows.  In
Section~\ref{sec:model} we very briefly summarize the model, with a
full description included in Appendix~\ref{app}. In
Section~\ref{sec:data} we present our data analysis and fitting
techniques, the results of which will be presented in
Section~\ref{sec:results}.  We discuss our conclusions in
Section~\ref{sec:discuss}.

\section{Model Details}\label{sec:model}
\subsection{Jet Model Direct Emission}

Our model for the jets in HS XRBs is based upon five main assumptions:
1) the total power in the jet scales with the total accretion power at
the inner edge of the disk, $\dot{M}c^2$, 2) the jet is expanding
freely and is only weakly accelerated via the resulting pressure
gradient, 3) the jet contains cold protons which carry most of the
kinetic energy while leptons do most of the radiating, 4) particles
are eventually accelerated into a power-law distribution, and 5) this
power-law is maintained along the length of the jet thereafter.  These
assumptions are motivated by observations which are discussed further
in Appendix~\ref{app}, where we have also included a detailed
description of the model and its development.  In this section we only
summarize the key parameters and assumptions of the jet model.

Beyond a small nozzle region of constant radius, which establishes the
base of the jet, the jet expands sideways with the initial proper
sound speed for a relativistic electron/proton plasma, $\gamma_{\rm
s}\beta_{\rm s}c\sim0.4c$.  The plasma is weakly accelerated along the
resulting longitudinal pressure gradient, which allows an exact
solution for the velocity profile via the Euler equation \citep[see,
e.g.,][]{Falcke1996}.  This results in a roughly logarithmic
dependence of velocity upon distance $z$.  After a period of more
rapid acceleration immediately beyond the nozzle, the velocity
gradient lessens, and the velocity saturates at Lorentz factors of
$\Gamma_{\rm j}\gtrsim$2-3.  The size of the base of the jet, $r_0$,
is a free parameter and once fixed determines the radius as a function
of distance along the jet, $r(z)$.

The the model is most sensitive to the fitted parameter $N_{\rm j}$,
which acts as a normalization.  It dictates the power initially given
to the particles and magnetic field at the base of the jet, and is
expressed in terms of a fraction of the Eddington luminosity $L_{\rm
Edd}=1.25\times10^{38} M_{\rm bh,\odot}$ erg s$^{-1}$.  The total
power input at the base of the jets is in fact approximately an order
of magnitude larger than $N_{\rm j} L_{\rm Edd}$, due to the
requirement of distributed acceleration along the jets (see
Appendix). Once $N_{\rm j}$ is specified and conservation is assumed,
the macroscopic physical parameters along the jet are determined.  We
assume that the jet power is evenly shared between the internal and
external pressures, and that the radiating particles are close to
equipartition with the magnetic field.  The radiating particles enter
the base of the jet where the bulk velocities are lowest, with a
quasi-thermal distribution.  Around $10-100$ $r_{\rm g}$, a
significant fraction of the particles are accelerated into a power-law
tail whose maximum energy is determined self-consistently with the
local cooling rates.

The particles in the jet radiatively cool via adiabatic expansion, the
synchrotron process, and inverse Compton upscattering; however,
adiabatic expansion is assumed to dominate the observed effects of
cooling because of distributed acceleration.  While thermal photons
from the accretion disk are included as seed photons in our Compton
calculations, the beaming reduces their energy density compared to the
rest frame synchrotron photons (synchrotron self-Compton; SSC), except
at the very base of the jet where they can be of the same order.
Reprocessed disk radiation will contribute even less and thus its
feedback on the X-ray spectrum will be negligible.  We therefore do
not include this latter component in our calculations.  In the case of
near-equipartition of energy between the magnetic field and particles,
inverse Compton processes only will dominate the direct synchrotron
emission close to the compact object where relativistic beaming is at
its minimum.

\subsection{Accretion Disk Feedback Signatures}

As the jet plasma is only weakly beamed, some fraction of the
radiation will impact upon the cooler material in the accretion disk
and lead to fluorescent line emission and a reflection hump
\citep[e.g.,][]{LightmanWhite1988}.  In \cite{MarkoffNowak2004} we
calculated the resulting reflection spectrum for typical jet models,
and found that when synchrotron emission from the acceleration region
dominates the hard X-ray spectrum, the fraction of reflected emission
is very small ($\sim$few percent).  In contrast, when Compton
processes (predominantly SSC) originating near the base dominate the
hard X-rays, `reflection fractions' of $\lesssim 20\%$ are possible.

As discussed by \cite{MarkoffNowak2004}, calculating the reflection
spectrum for our jet models is complex, as one needs to calculate a
reflected spectrum associated with each emission region along the axis
of the jet.  Furthermore, for our detailed jet models the disk sees a
qualitatively different jet spectrum than the direct spectrum viewed
by a distant observer.  This is in contrast to the simple `moving
corona' models of, e.g., \cite{Beloborodov1999}, where both the direct
spectrum and the spectrum impinging upon the disk have the same simple
power law shape, and the total reflected spectrum can be calculated
analytically.  For the initial spectral fit studies presented here, a
detailed reflection calculation for our jet model was deemed too
computationally prohibitive.

For the purposes of the study presented here, we adopt a simplified
approach.  In our fits, we add a single Gaussian line to our continuum
model, and we allow the line energy to vary between 6--7 keV and the
line width, $\sigma$, to vary between 0--1.5\,keV.  We then convolve
the entire \emph{directly viewed} spectrum with a non-relativistic
reflection model derived from the Green's functions of
\citet{Magdziarz1995}.  The amplitude of the reflected spectrum is
left as a fit parameter, expressed in the usual manner as a fractional
solid angle, $\Omega/2\pi$, subtended by the reflector.
($\Omega/2\pi=1$ indicates that the radiation impinging upon the
reflector is equal to that directly viewed by the distant observer.)
This general approach of phenomenologically fitting a reflection
spectrum is similar to that employed by many pure Comptonization
models \citep[e.g.][]{Coppi1999,Poutanen1998}, albeit here it is less
likely that the directly viewed spectrum is a completely adequate
proxy for the spectrum viewed by the cold disk.  We will consider more
sophisticated approaches to calculating the reflection features in
future works.

\subsection{Fitting Method}

In our previous papers
\citep[e.g.,][]{MarkoffFalckeFender2001,Markoffetal2003}, we only
compared the jet model to spectra that had been `unfolded' with
\texttt{XSPEC} using simple exponentially cutoff, broken power-law
models.  This is in fact the usual practice among many researchers
when comparing complicated theoretical models to multiwavelength data
\citep[e.g.,][]{EsinMcClintockNarayan1997,MarkoffFalckeFender2001}.
Unfortunately, such a procedure does not allow the determination of a
statistical goodness-of-fit, and more importantly it does not allow
one to compare the model to the fine features of the X-ray spectral
density distribution (SED).  Here we address this issue by
forward-folding the jet model through the detector response matrices
of the X-ray instrument, and then comparing to the data in `detector
space'.

Specifically, we have imported our jet + multi-color blackbody thermal
disk \citep[{similar to \tt diskbb};][]{Mitsudaetal1984,Makishimaetal1986} model as
a subroutine for use in standard X-ray data analysis packages.  The
current routine works in both {\tt XSPEC} \citep{Arnaud1996} and {\tt
ISIS} \citep{houck:00a}.  In this work, we perform our fits using {\tt
ISIS} v.1.2.6 as this latter package can read in and fit lower
frequency simultaneous data sets (e.g., radio through optical data) as
ASCII files, without the need for creating dummy response matrices.
Furthermore, the {\tt ISIS} `unfolded spectra' (shown in the figures
throughout this work) are independent of the assumed spectral model
(i.e., the unfolding is done solely with the response matrix and
effective area files; see \citealt{Nowaketal2005} for details).
All shown residuals, however, are for comparisons between the
properly forward-folded model and the data in detector space.  See
\cite{Nowaketal2005} for a discussion of further differences between
\texttt{ISIS} and {\tt XSPEC}.

We are making detailed comparisons between our jet model and detector
space data for the first time.  Indeed, to our knowledge, this is the
first time that such a complex, multi-wavelength model has been
compared properly to the X-ray data in detector space.  For this
paper, we therefore have chosen to initially leave most jet model
parameters free so that we can fully explore their effect on the fits.
We can then determine which parameters are most fundamental for
describing the data and which can effectively be frozen in future
applications.  This is essentially in the same spirit as prior studies
with Comptonization models, where models such as \texttt{eqpair}
\citep{Coppi1999} have significantly more parameters than are used in
a typical fit.  For such Comptonization models, even fairly basic
parameters, such as thermal vs. nonthermal compactness (i.e., the
ratio of coronal energy to coronal radius), can be degenerate with one
another in real data fits, and one often chooses to freeze the
nonthermal parameters to negligible values.  For a detailed discussion
of these points, see the review article by \citet{Coppi2004}.

Currently, our multi-component model runs significantly slower than
single-component models which calculate Comptonization in the corona.
Determining meaningful error bars for all parameters was not always
possible because for some fit parameters the resulting spectrum can
vary noncontinuously.  Given that this is the first time that
broadband data are being properly compared to our forward-folded
model, such initial difficulties are not surprising.  Nevertheless,
the model fits presented here represent an exploration of a
significant amount of the parameter space, especially given the
fitting time scales involved.

\section{Data}\label{sec:data}

We model the data from Rossi X-ray Timing Explorer (RXTE) observations
for the hard state Galactic black hole candidates (BHCs) GX~339$-$4
and Cygnus~X-1.  We use data from both the RXTE Proportional Counter
Array (PCA; \citealt{Jahodaetal1996}) and from the High Energy X-ray
Timing Experiment (HEXTE; \citealt{rothschild:98a}).  The data have
been extracted with the recent \texttt{HEASOFT 5.3.1} software
release. Compared to earlier releases of this software, the relative
calibration of the PCA and the HEXTE is now in excellent agreement
\citep[see analysis in, e.g.,][]{Wilmsetal2005}. Power-law fits
to spectra of the Crab pulsar and nebula show the PCA data to have a
systematic uncertainty of 0.5\%, which we added in quadrature to the
data.  We used the energy bands 3--22\,keV for the PCA (and grouped
the data to a minimum of 30 counts/bin) and 18--200\,keV for the HEXTE
(where we added data from the two clusters and grouped the counts to a
minimum signal-to-noise, after background subtraction, of 10 in each
bin).  We use the top layer of the PCA only. During part of the
observations, some proportional counter units of the PCA were switched
off. In these cases we added the spectra extracted for the different
PCU combinations, appropriately weighting the response matrices.

The Cygnus~X-1 data consist of three of the many observations taken
during our several years long RXTE monitoring campaign \cite[][and
references
therein]{Pottschmidtetal2003,Gleissneretal2004,Gleissneretal2004a} and
were chosen to be representative spectra bracketing the typical hard
state spectral variations of Cyg~X-1. The data from GX~339$-$4 also
have been discussed by us elsewhere
\citep{Wilmsetal1999,NowakWilmsDove2002,Nowaketal2005}, and bracket
the typical variations in the overall hardness and luminosity of the
\gx\ hard state.  (They do not, however, encompass the most extreme
bright and relatively soft HS data discussed by
\citealt{Homanetal2005},\citealt{Nowaketal2005}, and \citealt{Bellonietal2005}.)
Table~\ref{tab:obslog} contains a log of the observations.

\begin{center}
\begin{deluxetable}{cclcc}
\tablewidth{0pt}
\tablecaption{Log of Cyg~X-1 and GX~339$-$4 RXTE Observations \label{tab:obslog}}
\tablehead{\colhead{Source} & {Obs ID} & \colhead{Date} & \colhead{Exposure} & 
           \colhead{PCU off} \\
           & & & (sec.)}
\startdata
Cyg~X-1 & 40099-01-19 & 1999 September 25 & \phn8080 & 13 \\
Cyg~X-1 & 60090-01-26 & 2003 February  23 & \phn9696 & 1/4 \\
Cyg~X-1 & 60090-01-41 & 2003 September 22 & 11600 & 13/14 \\
\hline
GX 339$-$4 & 20181-01-02 & 1997 February 10 & 10528 & \nodata \\
GX 339$-$4 & 40108-02-01 & 1999 April 2 & 9152 & \nodata \\
GX 339$-$4 & 40108-02-03 & 1999 May 14 & 9776 & \nodata \\
\enddata
\tablecomments{The table lists those
  proportional counter units (PCUs) which were off during (part of)
  the observation. A slash denotes a logical ``or''.}
\end{deluxetable}
\end{center}

In order to compare our jet models to the more standard Comptonization
models, we also fit our observations with the hybrid thermal/non-thermal
Comptonization model \texttt{eqpair} \citep{Coppi1999}.  We slightly
changed the publicly available version of this code to make the model
run under {\tt XSPEC v.11.3.1}, and hence also under {\tt ISIS
v.1.2.6}. To allow for a partial covering of the disk by the Compton
corona, we add emission from a disk using the {\tt diskpn} model
\citep{Gierlinskietal1999}.  The peak temperature of this disk model is
also set equal to the peak temperature of the seed photons input to the
Compton corona. These seed photons are similarly assumed to have a
spectral energy distribution that follows that of the {\tt diskpn}
model.  The corona is presumed to be comprised solely of thermal
electrons, and its properties are described by two parameters: a seed
electron scattering optical depth, $\tau_{p}$ (pair production can
increase the total optical depth from this value, although pair
production is negligible for the fits described in this paper), and
the coronal compactness (i.e., coronal energy content divided by
coronal radius), expressed as a ratio relative to the soft, seed
photon compactness, $\ell_h/\ell_s$.

We also allow for an additive Gaussian Fe K$\alpha$ emission line
whose centroid energy is constrained to be between 6 and 7\,keV and
whose width is constrained to be $\sigma < 1.5$\,keV.  In our use of
the {\tt eqpair} model, the reflection component is relativistically
smeared using the velocity profile of the \texttt{diskpn} disk model,
which in turn follows the rotational velocity of a standard thin disk
circularly rotating under the influence of a Paczy\'nski-Wita
pseudo-Newtonian potential \citep{PW1980}.  The reflecting medium is
assumed to have solar abundances; however, the reflecting medium can
be ionized, as parameterized by the ionization parameter, $\xi$
\citep{done:92a}.

Table~\ref{tab:eqpair} gives the best fit parameters of these
\texttt{eqpair} fits, and the fits are shown in the third column of
Figs.~\ref{gx339_all} \& \ref{cyg_all}.  The reduced $\chi^2$ values
are close to unity, indicating very good agreement between the model
and the data. The parameters found for the Comptonization model are in
general agreement with those reported for earlier observations.  The
fits yield moderate optical electron depths of $\tau_{\rm es} \sim
1.5-3$ with coronal compactnesses of $\ell_h/\ell_s \sim 3.5$--$8$,
which correspond to coronal temperatures of $\sim 25$-100\,keV. (Note
that the $\tau_{\rm es} \sim 6.7$ for Obs ID 40108-02-03 has rather
large error bars, and is partly being driven by the poor statistics in
the HEXTE band.)

All of the coronal model fits yield reflection covering fractions
$\lesssim 20\%$. None of the fits strongly require ionized reflection;
however, the reflection model used by the {\tt eqpair} code is very
simplified (being based upon the {\tt pexriv} code;
\citealt{done:92a}).  The fitted equivalent widths of the Fe K$\alpha$
lines are on the order of 150\,eV, and consistent with the highly
smeared reflection models employed, the lines are generally found to
be rather broad. As will be discussed elsewhere in more detail
\citep{Wilmsetal2005}, we believe that part (although not all) of the
fitted line broadness is caused by the Comptonization model employed
here having too simplified a description of the transition between the
accretion disk emission and the Comptonization continuum.  Finally,
the changes in accretion disk temperature and flux, coupled with the
changes in coronal compactness and optical depth, are in response to
the fact that we have deliberately chosen observations that span a
wide range of flux and spectral hardnesses.

Detailed discussions of coronal model fits to Cyg~X-1 data (e.g.,
comparison to phenomenological broken power-law fits, comparison to
timing data) are discussed in \cite{Wilmsetal2005}.  Their are three
main points that we wish to emphasize with the fits presented here.
First, the {\tt eqpair} fits describe the data well -- as well as any
coronal models that we have explored (although not better than simple
exponentially cutoff, broken power-law models;
\citealt{Nowaketal2005,Wilmsetal2005}).  Second, the fits do
not require comparatively as large reflection fractions\footnote{The
{\tt pexriv} model, incorporated into the {\tt eqpair} model, tends to
have an unnaturally sharp ionized Fe edge structure, especially when
compared to the more sophisticated models of \citet{RossFabian2005}.
The reflection fraction can be artificially depressed when employing
the {\tt pexriv} model, so as to minimize residuals from its sharp
ionization edge.  However, as we employ a large degree of relativistic
smearing, and hence smooth out the edge, we do not expect to be
subject to this systematic effect.}.  Third, and related to the
previous point, the coronal models themselves describe a continuum
spectrum more complicated than a simple, exponentially cutoff
power-law.  Some fraction of the spectral hardening above 10\,keV,
normally attributed to the `reflection hump', is in fact described by
curvature of the Comptonization continuum.  In Compton coronal models,
this curvature is in part due to transiting from a spectrum strongly
affected by the soft seed photons to one strongly dominated by photons
that have been multiply Compton scattered. As we discuss below, this
continuum curvature an additional interpretation in jet models.

\section{Jet Model Results}\label{sec:results}

As mentioned above, our strategy in testing our jet models against the
data was to allow the maximum number of model parameters to vary
freely during the fit.  This allowed us to explore which parameters
had the greatest influence on the fits.  Several parameters settled on
fairly similar values for both sources and all observations, which
suggests that those parameters could be fixed in future applications
of this model. We will discuss this possibility explicitly below.

Fits with a complicated model can easily lead to false, local minima;
therefore, we began the fitting procedure outside of {\tt ISIS}, using
unfolded data sets in order to find a set of starting parameters that
would yield reduced $\chi^2\lesssim5$.  For this paper we have chosen
to explore only models with rough equipartition between the magnetic
and radiating particle pressures ($k\approx1$).  This assumption is
obviously not applicable for Poynting flux dominated jets; however, we
consider the very weak acceleration required by observations to be an
indication that magnetic domination (which would generally imply
stronger acceleration mechanisms) is not likely to be very extreme.

We use the results of \cite{MarkoffNowak2004} as a rough guide for
consistency, and only consider models with reflection fractions
$\lesssim 20\%$ for jet nozzle radii $r_0\lesssim20 r_{\rm g}$ to be
`successful' fits.  For smaller values of the radii, slightly lower
values for the reflection fraction would be expected.  In all cases
the jet models seemed to naturally prefer reflection fractions
$\Omega/2\pi \lesssim 20\%$, so in practice this restriction on
reflection fraction was not an issue.

\begin{figure*}
\epsscale{1}
\plotone{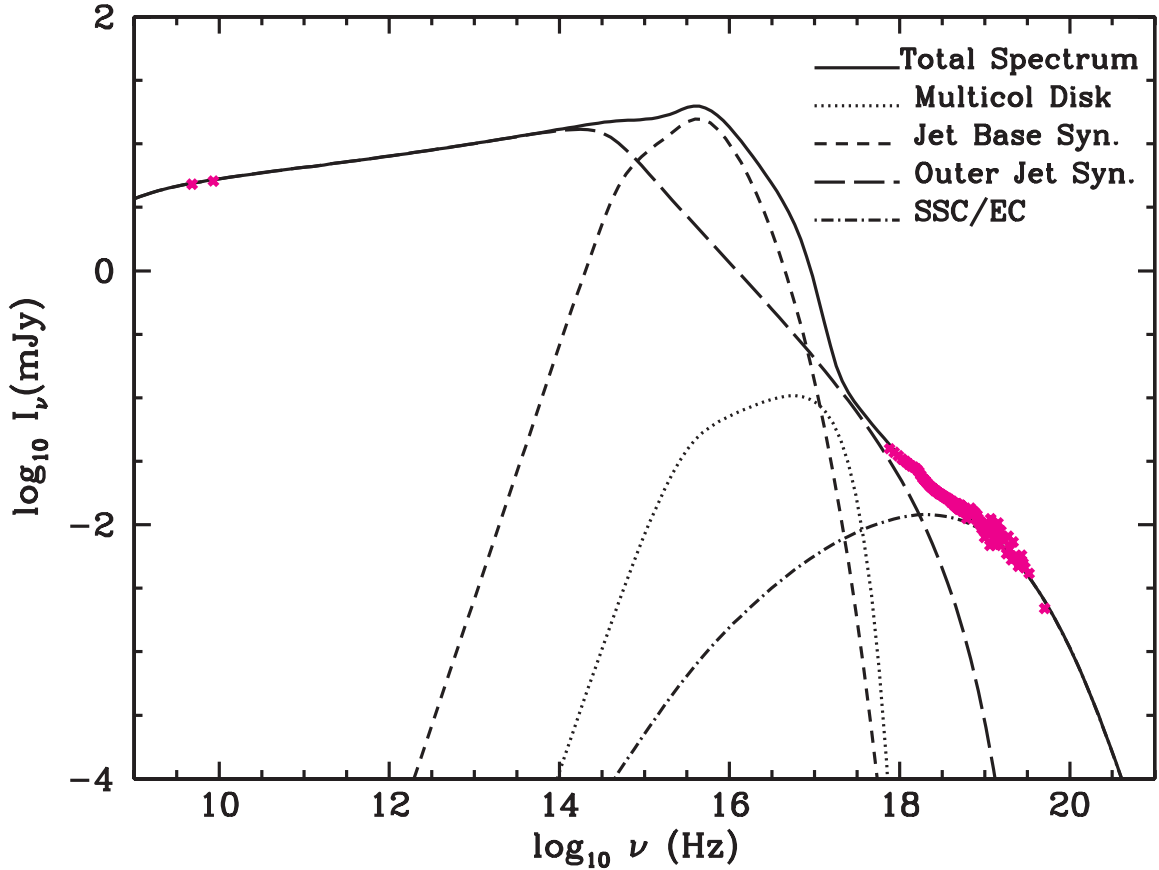}
\caption{Representative model for \gx\ obs.id. 40108-02-01, showing
  jet and multi-color blackbody models only.  This is the model which
  is then convolved with the disk reflection and line emission
  elements for the final statistical fit.  The components are
  labeled.  \label{gx339_comps}}
\end{figure*}

Fig.~\ref{gx339_comps} shows a representative model for \gx\ in order
to illustrate the contribution from the various components which go
into the later figures showing actual fits.  The radio through IR
originates exclusively from self-absorbed synchrotron radiation
beginning at $z_{\rm acc}$ and continuing outwards along the
jet. Optically thin synchrotron emission from the accelerated
power-law tail of leptons also contributes to the soft X-ray band.
The base of the jet radiates direct synchrotron photons, giving a
slight hump in the optical/UV, which are then included as seed photons
(along with disk photons) for upscattering by the emitting electrons
into an SSC/EC ``hump'' in the hard X-rays.  The shape of this hump is
rounded, due to the quasi-thermal particle distribution assumed in the
base, and thus reduces the need for a large fraction of disk-reflected
photons to contribute to the spectral break/hardening at $\sim$10 keV.
The possibility that there are two correlated continuum components in
the X-ray band is supported by our ensemble of simultaneous
radio/X-ray observations of Cyg X-1, whose spectra are very
well-described by exponentially cut-off, broken power-laws
(\citealt{Nowaketal2005,Wilmsetal2005}).  In these models, the soft
X-ray spectral slope is very well-correlated with the break of the
hard X-ray spectral slope, and the hard X-rays are very
well-correlated with the radio
\citep{Gleissneretal2004,Nowaketal2005}.  The multicolor blackbody
included in the jet model is also shown in the soft X-ray band.  The
photons from this component are included in the inverse Compton
upscattering within the jet.


\begin{figure*}
\epsscale{1}
\plotone{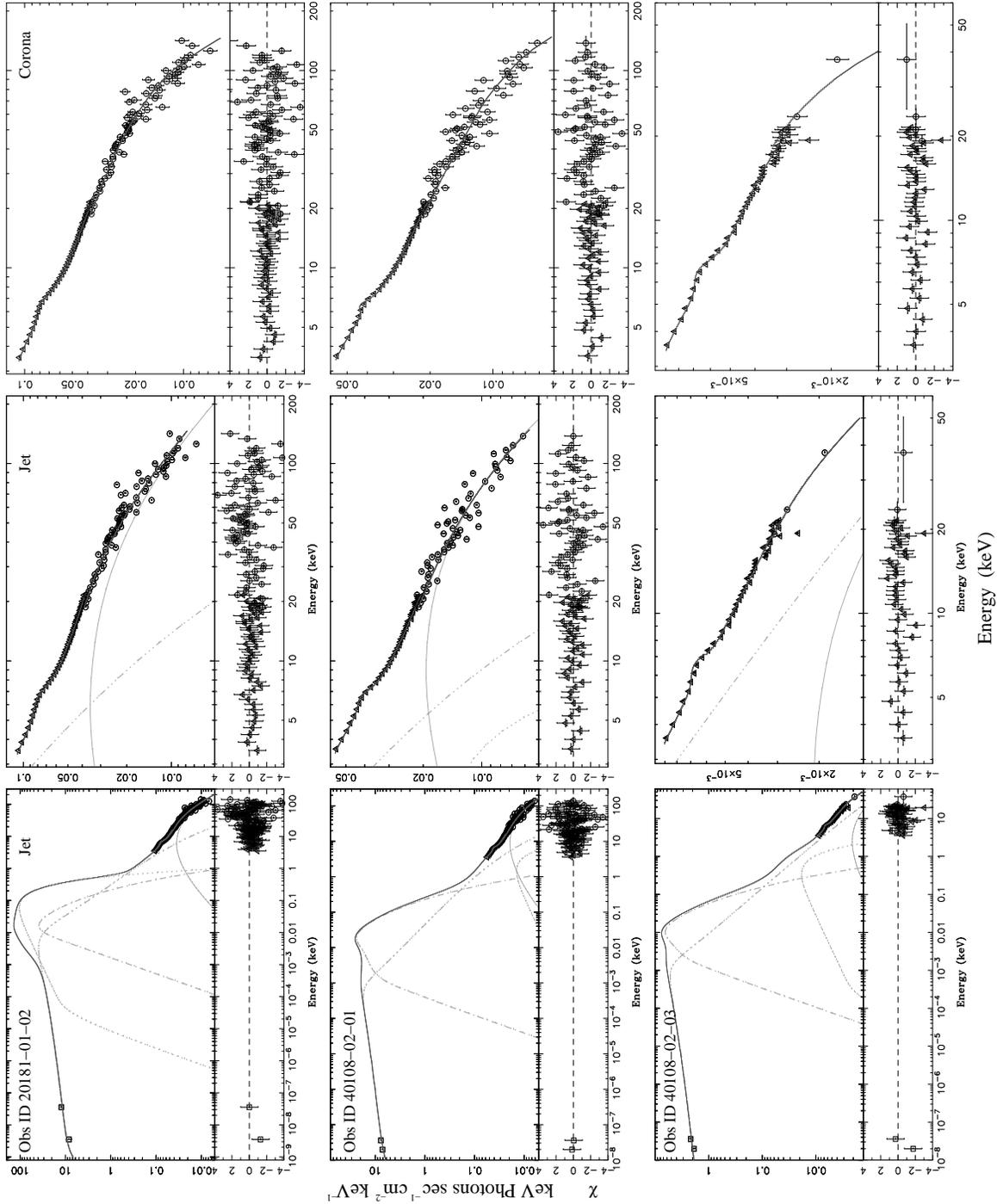}
\caption{Jet and corona model fits to radio and X-ray data of
GX339$-$4, with individual spectral components for the jet model shown
in grey.  In each panel, synchrotron from the jet pre-shock region is
shown as a dot-dash line, synchrotron from the post-shock region is
shown as a dash-triple dot line, synchrotron self-Compton plus
external Compton is shown as a solid line, and the disk radiation is
shown as a dotted line.  The panels are laid out as follows.  From
bottom to top (left to right on the figure), the fits are with the
{\tt mffjet} model (radio/X-ray and X-ray data) and then the {\tt
eqpair} model. From left to right (top to bottom on the figure), the
ObsIds are 20181-01-02, 40108-02-01, and 40108-02-03.  The spectra are
shown as model-independent unfolded spectra; however, the fits and
residuals are from the properly forward-folded models.
\label{gx339_all}}
\end{figure*}

\begin{figure*}
\epsscale{1}
\plotone{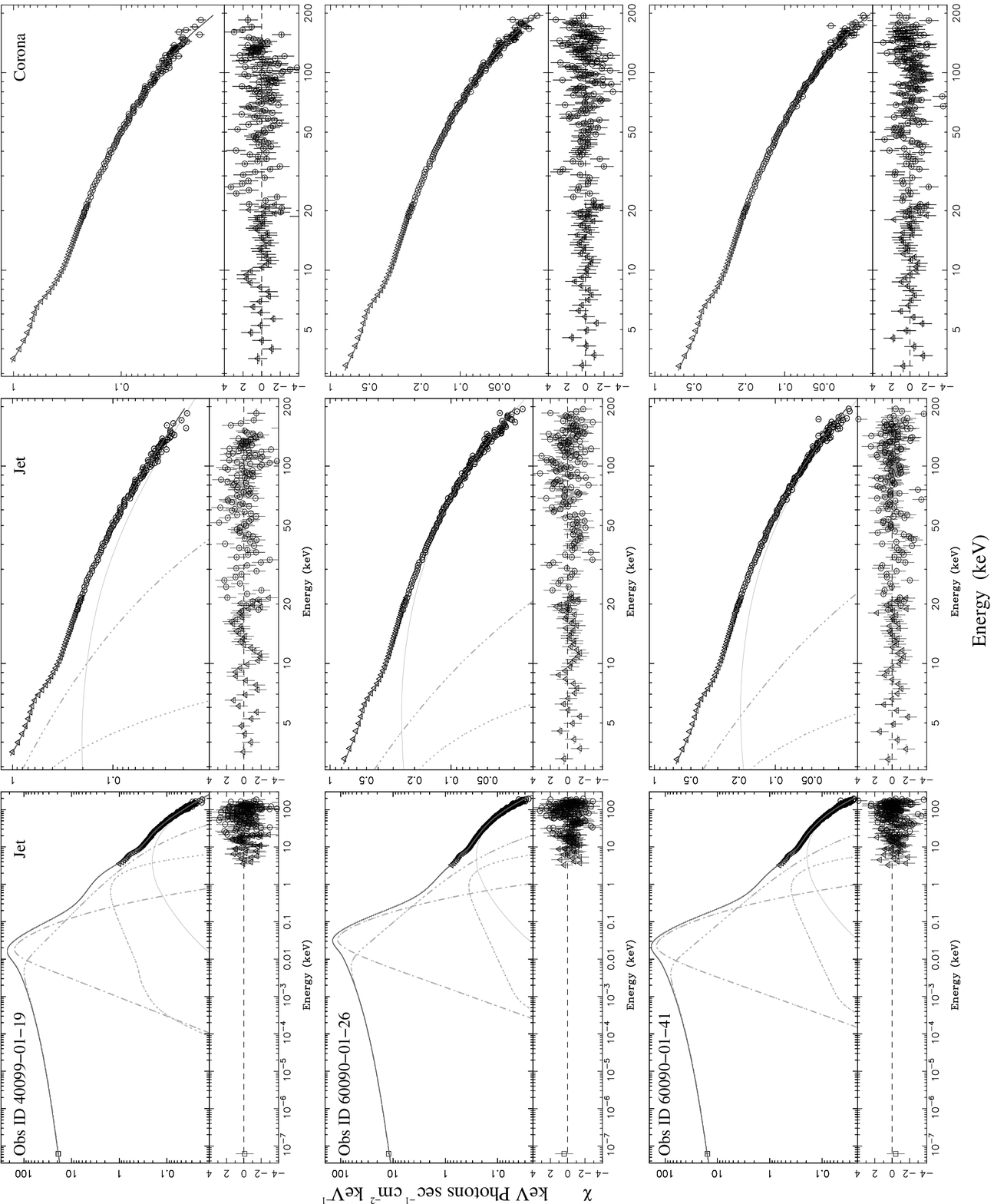}
\caption{Same as Fig.~\protect{\ref{gx339_all}}, except for
Cyg~X-1 data.  From left to right, (top to bottom on the figure), the
ObsIds are 4099-01-19, 60090-01-26, and 60090-01-41. \label{cyg_all}}
\end{figure*}

The statistical fits are shown in Figs.~\ref{gx339_all} \&
\ref{cyg_all}, for \gx\ and Cyg X-1, respectively.  Each figure
encompasses nine panels, with each row representing a single
simultaneous radio and X-ray observation of the source.  The first
column shows the entire radio through X-ray jet model + soft disk +
reflection fit, and the second column focuses just on the X-ray band.
The last column shows the thermal Compton corona model from {\tt
eqpair} for comparison.  Values for all fitted parameters are given in
Tables 2--4.

The results presented in Figs.~\ref{gx339_all} \& \ref{cyg_all} and
Tables \ref{tab:mffjetI}, \ref{tab:mffjetII}, \& \ref{tab:eqpair}
immediately demonstrate three important results of this work.
\begin{itemize}
\item {\bf Jet models describe the data equally well as pure Compton
  corona models, even when employing the broad-band and high
  statistics of the RXTE data.}  There has been question in the
  literature (e.g., \citealt{Zdziarskietal2004}) as to whether or not
  jet models can adequately describe the spectral cutoff of the hard
  tail. While pure synchrotron jet models may not be able to describe
  the steep cutoff present in some (but not all) observations, jets
  can also successfully account for this hard tail cutoff via
  Comptonization.  The primary differences between `traditional'
  coronal models and jet coronal models are that the coronal
  temperatures are higher in the latter ($\sim 10^{10}$\,K vs. $\sim
  10^{9}$\,K), and the disk seed photons in jet models are
  significantly augmented---if not dominated---by synchrotron seed
  photons.
\item {\bf The jet models, similar to the Compton corona models,
  describe a complex continuum with curvature -- specifically, a
  hardening above 10\,keV.}  Again, the hardening often predominantly
  ascribed to a `reflection hump' is being somewhat subsumed by
  continuum emission.  In the jet model, the steeper slope of the soft
  X-rays is primarily due to synchrotron emission from the jet, rather
  than being influenced by the disk photons, while the hardening is
  due to the SSC/EC component.  Reflection is present, but represents a
  smaller fraction of the total hard flux compared to the corona
  models.  Note that the fits discussed here are in contrast to
  earlier incarnations of the jet model which attempted to describe
  observations solely via the synchrotron component (e.g., the studies
  of XTE~J1118+480; \citealt{MarkoffFalckeFender2001}).  For
  XTE~J1118+480, the lack of any discernible break or hardening near
  10\,keV in the continuum spectrum was used to argue for the lack of
  a reflection component \citep{Milleretal2002}, which would be
  consistent with a jet spectrum dominated by the synchrotron
  component \citep{MarkoffFalckeFender2001,MarkoffNowak2004}.  In
  contrast, the fact that all of the XRB spectra shown here harden
  above 10\,keV argue for the importance of both the synchrotron and
  SSC/EC components of the jet spectrum for these fits, as well as the
  presence of reflection.

\item {\bf Although the jet and Compton coronal models describe the
  X-ray data equally well, only the jet model naturally describes both
  the amplitude and slope of the radio data without the need for
  additional free parameters.} It is important to note that the radio
  spectrum, as well as the bulk of the 3-200\,keV X-ray continuum
  (i.e., the overall amplitude and continuum curvature), are almost
  solely driven by the parameters of the jet: energy input to the jet
  ($N_{\rm j}L_{\rm Edd}$), radius of the jet base ($r_0$), electron
  temperature at the jet base ($T_e$), slope of the power-law
  component of the electron energy distribution ($p$), the
  equipartition parameter ($k$), and the location of the particle
  acceleration zone in the jet ($z_{acc}$).  If one begins instead
  with a corona model and attempts to fit the amplitude and slope of
  the radio data by assuming a relationship to an extant jet model,
  this will in essence require the addition of many more parameters to
  describe just two physical quantities.  We argue that nothing can be
  gained by this approach, which is why we have chosen to study the
  test case of full coupling.
\end{itemize}

From our set of jet model fits that successfully describe the
GX~339$-$4 and Cyg~X-1 data, we can now explore the implications of
the fitted parameters.  Cyg X-1 and \gx\ are somewhat different
sources while in their hard states.  The luminosity of \gx\ varies
much more than that of Cyg~X-1. This is apparent in the fitted values
for the power normalization $N_{\rm j}$ and output power $L_{\rm j}$
shown in the first two columns of
Table~\ref{tab:mffjetI}.  Both sources show best fit $N_{\rm j}$  in the
range of $10^{-4}-10^{-3}$ $L_{\rm Edd}$, but \gx\ shows significantly
more variation in this parameter.  Differences can also be seen in the
other main parameters, for instance on average \gx\ favors slightly
larger values of the nozzle radius, $r_0$.  This reflects the slightly
higher ratio of X-ray to radio flux observed in Cyg~X-1.  A smaller
scale jet base, for a given jet power, increases the jet compactness
which in turn gives a slightly higher flux (with the X-ray flux being
more sensitive to this effect) and pushes all jet emission to slightly
higher frequencies.

Other fit parameters also indicate differences between these two
sources; however, the parameters surprisingly fall roughly in the same
range given the \emph{a priori} possibility for much greater parameter
variations.  Note that the lowest luminosity \gx\ observation, Obs ID
40108-02-03, has poor enough statistics that the following general
statements about the \gx\ fits will be based mostly on the brighter
two observations.  For example, both the electron temperature, $T_e$,
and the location of the start of the acceleration zone, $z_{\rm acc}$,
tend to be larger in GX~339$-$4 than in Cyg~X-1.  (Although we were
unable to find adequate error bars for the latter parameter, the trend
of finding larger values in \gx\ was persistent.)  Additionally, we
note that the electron power law index, $p$, shows more variation in
\gx\ than in Cyg~X-1.

Although the range in electron temperatures at the base of the jets
falls within a range of a factor of $<2$ for both sources, one of the
criticisms of this class of models has been that $\sim
3$-$5\times10^{10}$ K is not as ``natural'' a value as the $\sim
100$\,keV typically used in thermal Comptonization models.  The jet
model electron temperature, however, is comparable to or slightly
greater than the value typically derived for radiatively inefficient
accretion flows \citep[e.g.,][]{Narayanetal1998}.  If some fraction of
the accreting plasma is fed directly into the jets, and also perhaps
heated slightly in the process, we would expect such a temperature.

On the other hand, several jet model free parameters seemed to settle
quite quickly into similar values for both Cyg X-1 and \gx: the
equipartition factor $k$, the fraction of accelerated particles
$pl_f$, and the ratio of nozzle length to radius $h_0$.  Physically,
this suggests that these jets either are close to equipartition, or we
have found a local minima of solutions for $k\approx1-2$, and we may
find other reasonable minima for $k\gg1$, e.g., magnetically-dominated
jets.  We plan to explore this particular question in a future work.
It seems reasonable, however, to freeze the fraction of accelerated
particles at $\sim 75\%$ and make the statement that reasonably
efficient acceleration is expected to occur in hard state BHC jets.
Similarly, a compact base/corona region with scale height similar to
the radius seems a reasonable assumption.  It is likely that this
parameter can also be frozen in future applications of the jet model.

We listed the two ``acceleration parameters'', the shock speed
relative to the bulk plasma flow, $u_{\rm acc}/c$, and the ratio of
the scattering mean free path to the gyroradius, $f_{sc}$, as separate
fit parameters.  However, as discussed in the Appendix, these
parameters are perhaps not physically meaningful as currently defined,
since we are no longer as convinced that the acceleration process is
diffusive Fermi acceleration.  These parameters enter into the
acceleration rate as the factor $f=(u_{\rm acc}/c)^2/{f_{sc}}$, which
is compared to the sum of the cooling rates from synchrotron and
inverse Compton radiation, and adiabatic expansion in order to
calculate the local maximum accelerated lepton energy.  Therefore,
this parameter can be loosely interpreted as a factor related to the
efficiency of the acceleration process and merged for future fits into
a single parameter.  For the current fits, the individual components
were left free to vary, with mixed results.  For Cyg X-1,
$f\approx2-3\times10^{-4}$, suggesting a meaningful range of fits.
However, for \gx, $f$ ranges over two orders of magnitude, and we
suspect that we did not fully explore parameter space meaningfully for
this source.  In future applications of the model, it is likely that
we will combine these two parameters into a single parameter that will
absorb our uncertainty about the acceleration process.

While we list best fit values for the multicolor blackbody disk model
parameters $L_{\rm disk}$ and $T_{\rm disk}$, as well as the derived
inner radius $r_{\rm in}$, it is important to keep in mind that
because the data only extend down to a few keV, we cannot meaningfully
constrain this component from spectral fitting.  As described in the
Appendix, the disk photons are not as important to the overall photon
field as the locally produced synchrotron photons and thus can mainly
be constrained by their direct spectral contribution.  This
contribution, although weak, is in fact required for a good fit, but
it is not unique.  \cite{Ibragimovetal2005} have also fit some of the
X-ray data presented here, and for similar reasons fixed the thermal
disk parameters.  On the other hand, for the disk emission plus
Comptonization models presented here (as well as for those presented
in \citealt{Wilmsetal2005}), the temperature of the disk was tied to
the temperature of the seed photons---i.e., the thermal photons were the
\textsl{only} source of seed photons for Comptonization.  Thus by
virtue of these imposed restrictions, much more stringent formal
limits for the disk components were attainable.  

The main result to take away about the accretion flow modeling is that
our fits are generally consistent with a sub-Eddington accretion disk
with temperatures somewhat less than 1 keV.  Similarly, the total
power entering into the jets is roughly consistent with being of the
same order as the observed luminosity required in the disk to be
consistent with the data.  The presence of a weak disk component is
necessary for a good fit, and thus disk photons will contribute to the
inverse Compton component from the jet.  We do not, however, feel we
can confidently make any statements about the disk geometry and thus
the accretion rate at the inner radius assumedly feeding the jets.

For both \gx\ and Cyg~X-1, the amount of reflection required by the
jet models is roughly comparable to, although in general slightly
lower than, that required by Comptonization models.  This trend can be
understood by the fact that the jet base SSC/EC component dominates above
10\;keV, and to some extent subsumes the role played by the ``Compton
hump'' normally attributed to disk reflection.  This effect of course
brings up an interesting point: there is a clear degeneracy in how the
spectral hardening above 10\;keV can be understood in terms of
continuum models.  The fact that the jet SSC/EC component has a similar
appearance to the Compton reflection hump does not preclude the
presence of both.  Clearly, the fluorescent Fe line implies that there
must be a degree of reflection.  What these results (as well as the
Comptonization fits; \citealt{Wilmsetal2005}) do suggest is that one
cannot uniquely determine a reflection fraction independently of the
presumed continuum model.

\section{Discussion and Conclusions}\label{sec:discuss}

The key result of this work is that we have clearly demonstrated that
even the high statistics of RXTE X-ray data cannot distinguish between
the jet model and thermal Comptonization. (Indeed, the RXTE data also
cannot distinguish between thermal and nonthermal Comptonization
models, as has been previously discussed by, e.g.,
\citealt{Coppi2004}).  In this sense, the base of the jets
can be said to effectively ``subsume'', at least spectrally, the role
created for the corona. The main difference between these two pictures
then comes down to geometry, inflow versus outflow, and the
relationship of the corona to the lower-frequency-radiating regions of
the outer jets.  The jet model presented here is consistent with the
physical picture seen in relativistic MHD simulations
\citep{StoneMiller2000,HawleyKrolik2001,MeierNakamura2004}, where the
corona is not static but instead is a windy hot material blowing away
from the inner regions of the accretion flow.  This wind/corona is
likely what enters into, or is collimated and becomes, the jets and
therefore it is intimately related to the larger scale outflows,
similar to what we propose here.  In addition, the jet base provides a
natural illuminator for the disk.  

Our ability to constrain the role of the accretion disk is
significantly limited by the lack of low-energy data.  Ideally we
would like to be able to determine the inner temperature, and perhaps
also assess the need for a recessed inner edge, explicitly from the
observations.  Having this information would allow us to better
constrain the ratio of external to synchrotron-produced photons in the
jet base.  Additionally, we expect differences between jet models and
Compton corona models to become more pronounced as one considers data
above the nominal Comptonization cutoff at 100--200\,keV.  From both
of these perspectives, the observational situation will likely soon
improve with the launch of ASTRO-E2, which will have the ability to
observe from 0.5-600 keV and thus more stringently constrain both jet
and Compton corona models.  Additionally, although the broad-band
effects of reflection, e.g., the reflection hump, cannot be fit
independently from the assumed continuum, there is some hope that
finer spectral resolution (perhaps coupled to correlated studies of
the continuum variability) will be able to more uniquely to determine
the reflection fraction based upon detailed modeling of the Fe line
\citep{YoungReynolds2000,RossFabian2005}.

In the scenario considered here, the nozzle/base of the jet is the
dominant region for the creation of the hard X-ray emission via
Comptonization processes.  Constructing a completely realistic
description of this region, however, is difficult given the that the
physics of jet formation, acceleration, and collimation are not yet
understood.  The nozzle essentially represents the initial conditions
which fix parameters along the rest of the jet.  We have shown that
the form we have developed for this region over the last several years
can successfully mimic the corona in purpose, but it is not as
well-determined feature as we would like.  An alternative, but also
currently rather ad-hoc, possibility would be to invoke special
conditions linking an outflowing corona with more typical thermal
plasma conditions to an outer jet.  This would require some kind of
acceleration and/or heating of the particles at the interface (and
thus more free parameters), but may satisfy many of the spectral
requirements.

One possible way to discern between these two situations, or at the
very least limit the possible flux from a nozzle, could be to
calculate the amount of local ionization expected from nozzle UV
photons and compare that to what is observed (J. Krolik, priv. comm.).
If it can be shown for a given source that all observed line emission
can be accounted for by the known companion, this limit may be rather
stringent.  We note, however, that in High Mass X-ray Binaries such as
Cyg~X-1, the companion luminosity, especially in the UV, is extremely
high.  It is therefore more likely that meaningful limits can be
obtained for Low Mass X-ray Binaries, such as \gx.

While we have focused on modeling simultaneous radio/X-ray data sets
in this paper, for \gx\ and several other sources observations do
exist which in addition have simultaneous IR/optical data.  Near-IR
data in particular can provide a valuable constraint on the jet
synchrotron component, especially when the transition turnover from
optically thick to thin is resolved \citep[e.g.][]{Homanetal2005}.
Using the work presented here as groundwork, we will explore elsewhere
whether more parameters can be constrained by such data sets.
Similarly, with {\em HESS} already online and {\em GLAST}
imminent, predictions of $\gamma$-ray fluxes will be another important
way of testing not only the current model, but models which
incorporate hadronic processes.  

However, the next major frontier for application of our jet model
clearly is to consider questions of time-dependence.  By studying the
lag between the radio and X-ray correlated variability, we can
constrain the plasma speed and/or distance between the two
regions. The existing radio/X-ray flux correlations constrain the long
time scale, while simultaneous radio/RXTE pointed observations
constrain thousand of second time scales \citep{Gleissneretal2004}.
We are currently using long duration (several hundred ksec),
quasi-continuous radio/X-ray observations of Cyg~X-1 to explore
intermediate time scales (Markoff et al., in prep.).  The models
considered here, however, are steady-state and focus on the hard
accretion state only.  Some of the most interesting and revealing
behavior is seen during transitions between these states.  For
instance, during transition into the hard state we observe a hard
X-ray tail before the jet forms as a detectable radio structure
\citep{NowakWilmsDove2002,Milleretal2004}.  The presence of the tail
suggests that the jet base/corona may form as a viable region before
the outer stable radio-emitting structures are built up.  There are
also significant clues coming from noise and faster scaled time
variability such as QPOs \cite{HomanBelloni2005} which need to be
integrated into the picture.  In general the state transitions reveal
many interesting properties which are clearly not due to a steady
state model, but which do share several similar properties to the more
persistent hard states.  Studying the time-dependent behavior of these
systems will likely be an effective way to further constrain the
conditions at the jet base/corona.


We acknowledge partial funding and travel grants from the Deutscher
Akademischer Austauschdienst (DAAD) and the National Science
Foundation, under NSF grant INT-0233441.  M.A.N. is also supported by
NASA Grant SV3-73016.  S.M. is supported by an NSF Astronomy \&
Astrophysics postdoctoral fellowship, under NSF Award AST-0201597.
We thank the Aspen Center for Physics for its hospitality during the
final stages of the preparation of this paper, as well as the 
anonymous referee for helpful comments.  

\appendix
\section{Detailed description of the jet model}\label{app}

\subsection{Background}

Like all models for optically thick outflows, our jet model builds on
the initial work of \cite{BlandfordKoenigl1979}.  Their paper
demonstrated how a superposition of idealized, self-absorbed, conical
jet components results in a flat synchrotron spectrum in the radio
bands.  In reality, compact jets usually deviate from this idealized
case, and show a slight spectral inversion in the radio wavebands,
with spectral index $\alpha\sim0.0-0.2$ ($F_\nu \propto \nu^\alpha$).
A more realistic treatment of the internal physics can account for
this inversion.  Using the framework for hydrodynamical jets developed
in \cite{FalckeBiermann1995}, \cite{Falcke1996} developed a ``free
jet'' model which is more self-consistent.  For instance, the
assumption that the jet is roughly conical via lateral expansion with
a constant sound speed requires weak longitudinal acceleration.  The
Euler equation can then be solved for the longitudinal velocity
gradient, assuming only adiabatic losses, and is analogous to a
pressure driven wind solution with the proper jet speed replacing the
wind speed.  This weak acceleration combined with relativistic beaming
effects will result in the spectral inversion typical of compact jet
sources.

These earlier models were very useful for studying the general
physical scalings predicted by compact jet models.  However, without
full particle distributions and radiative transfer, they cannot be
tested against observed spectral data.  Our models have been developed
on the basis of these earlier papers, but specifically focus on the
problem of spectral predictions.  These models are only appropriate
for systems where the jets are not expected to be highly collimated
nor highly accelerated, e.g., the jets in hard state XRBs and LLAGN.

\subsection{Jet Physical Parameters}

We start by assuming that the base of the jets are at the speed of
sound for a relativistic gas with adiabatic index 4/3, giving a proper
sound speed $\beta_{\rm s}\gamma_{\rm s}\sim0.4$.  Furthermore, we
assume the jets expands laterally with the constant sound speed, which
is appropriate under the assumption of maximally efficient jets
\cite[see][]{FalckeBiermann1995}.  The resulting longitudinal
acceleration along the axes is largest near the bases, and
asymptotically approaches a maximum of Lorenz factor $\gamma_{\rm j}\sim
2-3$ in the outer radio emitting regions, and $\gamma_{\rm j}\lesssim1.5$
for the synchrotron emission.  Such weak beaming factors are suggested
by observations of hard state jets
\citep{Maccarone2003,GalloFenderPooley2003}, in contrast to the near
light-speed velocities implied by superluminal motion in higher
luminosity radio-emitting states.

The main free parameters of this model are the power normalization
$N_{\rm j}$, the radius of the jet nozzle $r_0$, the equipartition of
energy between the radiating leptons and the magnetic field $k=U_{\rm
B}/U_{\rm e}$, and the parameters determining the initial lepton
distribution as described below.  We also left the length of the
nozzle region (with fixed radius), $h_0$ free, but it turns out the
fits are not very sensitive to its value.  Other parameters which come
in, such as the central mass $m_{\rm bh}$, the inclination of the jets
to the line of sight $\theta_{\rm i}$ and the distance to the source
$d_{\rm kpc}$, are determined from observations and remain fixed for
any given source.  Finally there are two parameters related to the
presence of weak disk emission, the total luminosity radiated by the
disk and its inner radius temperature, $L_{\rm d}$ and $T_{\rm in}$
respectively.  These cannot be well constrained by our data because
although they are spectrally fit assuming a multi-color disk blackbody
\citep{Mitsudaetal1984,Makishimaetal1986}, the low-energy X-ray
spectrum is not well covered in our observations.  Similarly, although
they do contribute to the photon field upscattered by energetic jet
leptons, they are not the dominate component.

The most important free parameter is $N_{\rm j}$, parameterized in
terms of the Eddington luminosity, which determines the power
initially input into the particles and magnetic field at the base of
the jets.  The total power input at the base of the jets is in fact
approximately an order of magnitude larger than $N_{\rm j} L_{\rm
Edd}$, and for our model can be estimated only \textsl{a posteriori}
from the resulting emission in the emitting rest frame.  The
difference between $N_{\rm j}L_{\rm Edd}$ and the total output
radiation $L_{\rm j}$ expresses the lack of knowledge of what
initially collimates the nozzle, as well as what causes the
distributed reenergization---and consequent energy requirement---to
maintain the observed \cite[e.g.][]{Jesteretal2001} particle
distributions along the jet (see below).  $N_{\rm j}$ plays a similar
role to the compactness parameter in thermal Comptonization models,
which parameterizes the total energetics in the absence of an
understanding of the mechanism for energizing the corona
\citep{DoveWilmsBegelman1997,Coppi1999},

As a consequence of assuming maximal jets, there should be approximate
equipartition between the kinetic and internal energy.  Once $N_{\rm
j}$ is fixed, therefore, the energetics along the jets are fixed as
well.  For a given $r_0$ and $k$ at the base, conservation of energy
and particle flux together with the proper velocity profile along the
jet, $\gamma_{\rm j}(z)\beta_{\rm j}(z)$, determine the radius,
density, and magnetic field profiles along the entire jets.  What is
left to be determined is how the energy in the radiating particles is
distributed.

\subsection{Radiating Particle Distributions}

As mentioned earlier, we assume that the jets contain both leptons and
hadrons, but that the hadrons remain non-relativistic and serve only
to carry the bulk kinetic energy.  The leptons, which are assumed to
be mildly relativistic and quasi-thermally distributed \cite[see,
  e.g.,][]{QuataertGruzinov1999} are thus more likely to be
accelerated.  In any case, we assume that leptons are predominantly
responsible for the observed radiation.  This is begging the question
of hadronic acceleration, because the end effect of significant
hadronic acceleration and subsequent inelastic collisions will
invariably be relativistic leptons.  The expected energy
distributions, whether from direct acceleration or creation via
hadronic collisions, are not trivial to discern from X-ray emission
alone.  Therefore we do not specify whether the radiating particles
are electrons or positrons, since this will not effect the observable
outcome.  Eventually the consistency with these results can be checked
against hadronic models, but this is a very difficult problem
\citep[see,
  e.g.][]{Mannheim1993,MarkoffMeliaSarcevic1999,BoettcherReimer2004}
and is not considered here.  The one hint for what will be required
beyond neutral matter is that we do not require $n_{\rm e}/n_{\rm
  p}=1$, and once a fit is made, we can see if additional pair
creation is necessary.

The current model was developed in simplified form by
\cite{FalckeMarkoff2000}, where we first incorporated simple particle
distributions for the radiating leptons and calculated the radiative
transfer along the jets.  Our goal was to determine if the same kind
of model which could explain the inverted radio core of the Galactic
Center supermassive black hole, Sgr A*, also predicted significant
X-ray emission.  This idea was motivated by the first identification
of Sgr A* in the X-ray band by the {\em Chandra} Observatory
\citep{Baganoffetal2003}.  We explored two ``canonical'' types of
particle distributions, power-laws and Maxwellians, for Sgr A* in its
quiescent state \citep[see, e.g.,][]{MeliaFalcke2001}.  These two
ostensibly different distributions result in similar fits as long as
the characteristic particle energy, ($\gamma_{\rm e,min}$ for the
power-law or $\gamma_{\rm e, peak}$ for the Maxwellian) is similar.
Unlike ``typical'' low-luminosity AGN \citep[LLAGN;][]{Ho1999}, the
quiescent Sgr A* does not show any indication of a power-law of
optically thin synchrotron emission after the break from its
flat/inverted radio spectrum.  Therefore if the radiating particles
have a power-law distribution, it must be so steep as to be
indistinguishable from a Maxwellian in the optically thin regime.
Either scenario strongly suggests, for example, that acceleration in
the jets of Sgr A* is absent or very inefficient at
$L\sim10^{-9}L_{\rm Edd}$.

Based on Sgr A*, we therefore assume that the baseline particle
distribution, below a certain accretion rate which must be very low,
is quasi-thermal.  This scenario is consistent with particles being
advected into the jets directly from the accretion flow, or created
and roughly thermalized near the base. Sources with jets at higher
luminosity, such as AGN and XRBs, do, however, show the standard
indicator of particle acceleration: optically thin synchrotron
emission from a power-law distribution of particles \citep[e.g.,][and
refs. therein]{MarscherGear1985,FenderKuulkers2001}.  Interestingly,
Sgr A* shows daily flaring in which the X-ray spectrum rapidly
increases in flux and hardens dramatically
\citep{Baganoffetal2001,Baganoff2003}.  The flare emission stems from
energization of the particles via acceleration, heating or both at
once \citep{Markoffetal2001}, resulting in enhanced synchrotron and/or
SSC emission in the X-rays.  Because the increased X-ray flux during
the flares brings Sgr A* more in line with typical levels of LLAGN
activity \citep{Markoff2005}, one interpretation is that acceleration
of particles is the main factor contributing to the comparative
weakness of Sgr A*.  We also therefore assume that above some critical
accretion rate above that of Sgr A*, particle acceleration will be
commonplace in the jets.  This is also necessary to explain the high
synchrotron efficiency in XRBs \citep[see, e.g.,][]{Markoffetal2003}.

In the only XRB where the optically thick-to-thin turnover from
compact jet synchrotron has been directly observed
\citep[\gx\; ;][]{CorbelFender2002,Homanetal2005}, it occurs at a
significantly lower frequency than would be expected if the
acceleration began in the jet base.  We therefore assume that when
particle acceleration occurs, it begins at some location $z_{\rm acc}$
along the jet, which is a free parameter but which appears to fall in
the range of a few $10-10^2$ $r_{\rm g}$ for most sources we have so
far considered.  The fraction of particles accelerated into the tail
is a free parameter in our model, but is generally driven to fairly
high values so we will likely fix it at some value above 80\% in
future applications.  The power-law index of the tail $p$
($N(E)\propto E^{-p}$) is also a free parameter, and can be constrained
by the data.

Particle acceleration must compete with radiative and adiabatic
cooling to energize the quasi-thermal particles into a power-law tail.
Where the sum of the cooling rates equals the acceleration rate
defines the maximum achievable particle energy, $\gamma_{\rm e, max}$.
Unfortunately since the acceleration process is still an open
question, the acceleration rate is open to significant interpretation.
In \cite{Markoffetal2001} and \cite{MarkoffFalckeFender2001}, we used
acceleration rates appropriate for diffusive shock acceleration in the
most conservative case of the magnetic field direction parallel to the
shock normal \citep[see][]{Jokipii1987}.  The rate of diffusive shock
acceleration is always proportional to the magnetic field.  If
synchrotron cooling dominates the cooling term, $\gamma_{\rm e, max}$
will thus be independent of the magnetic field and can give
information about the plasma parameters if the location of the cutoff
can be determined.  In \cite{MarkoffFalckeFender2001} we showed
that this results in a synchrotron emission cutoff around 100 keV for
$\beta_{\rm sh}\gtrsim0.5$ and if the ratio between the particle's
mean free path for diffusive scattering to the gyroradius,
$\xi=\lambda/r_{\rm B} \approx 100$.  In general $\xi$ is thought to be
limited by the ratio of the particle to shock velocities, and so
likely no higher than $10^2-10^3$, but this value is certainly not
fixed.

At the time of writing, other processes such as stochastic resonant
acceleration \citep[e.g.,][]{Miller1998} are coming into favor.  Given
this uncertainty in the mechanism, we are loosening the constraints on
the acceleration rate we used previously and are letting $\xi$ and
$\beta_{\rm sh}$ vary.  If the acceleration rate we are using is not
really physical, then these parameters have lost their meaning and
become essentially fudge factors absorbing the free parameters
controlling the acceleration rate for other mechanisms, to be
determined later.  If shock acceleration does still hold, then these
tell us about the plasma conditions in the jets.  

Regardless of the process, the inferred cooling times for the
accelerated leptons will be too high to maintain the power-law along
the jets unless there is continuous distributed acceleration.
Distributed, continuous \textsl{in situ} injection is a standard
requirement for explaining the persistence of particle distributions
along AGN jets \citep{Jesteretal2001} and seems to be a common feature
in outflows.

So to summarize the particle distribution discussion, we assume that
the particles enter the jet with a relativistic quasi-Maxwellian
distribution, the temperature of which, $T_{\rm e}$, is a fitted
parameter. Some fraction of the particles are then assumed to be
accelerated continuously beyond a location in the jets, $z_{\rm acc}$.
Where exactly can be constrained by the frequency of the
optically thick-to-thin break in the synchrotron spectrum, if
detected, as well as from the synchrotron component contributing to
the soft X-rays.  

\subsection{Jet Continuum Components}\label{sec:jetcomps}

As the leptons travel outwards along the jets, they interact with
the local magnetic field and photon fields and cool via synchrotron and
inverse Compton radiation.  Inverse Compton processes are strongest
near the base of the jets, where the density is highest, and thus one
X-ray component is due to the upscattering of both jet synchrotron
photons as well as thermal disk photons by the quasi-thermal leptons
in the base.  For the assumptions in this paper (the thin disk is
recessed with inner radius $r_{\rm in}$ calculated from fitted
parameters $L_{\rm d}$ and $T_{\rm in}$), the disk thermal photon field is
only an important factor at the very base of the jets. This is because
even the mild beaming in the jets quickly serves to weaken the disk
photon field's energy density compared to the locally produced
synchrotron photons.  

Once the acceleration zone is reached, it is difficult to suppress
X-ray synchrotron radiation from the tail of accelerated leptons in
the diffusive acceleration case
\citep[see][]{MarkoffFalckeFender2001}.  As described above, we are
relaxing the constraint on the synchrotron cutoff by allowing for an
alternate acceleration process.  In this scenario, a power-law of
synchrotron radiation will contribute significantly at least to the
soft X-ray band, but not necessarily to the hard X-rays which are in
this model dominated by inverse Comptonized disk and synchrotron
photons.  

This distinction is important because, as shown by
\cite{MarkoffNowak2004}, a very typical hard state model where the
X-rays are exclusively due to synchrotron radiation cannot easily
reproduce disk reflection fractions greater than a few percent
(assuming perpendicular geometry and no disk flaring).  The larger
distance of the first synchrotron emission region $z_{\rm acc}$ from
the disk, combined with moderate beaming, reduces its importance for
disk feedback.  On the other hand, inverse Compton processes near the
base of the jets can produce reflection fractions of up to $\sim
15$--20\%.  This value is actually a lower limit because it does not
include the effects of light bending, which will serve to increase the
reflection fraction possibly quite significantly
\citep{MiniuttiFabian2004}

Beyond relaxing the assumption about the acceleration rate, there are
a few other differences between this model and the
synchrotron-dominated models considered in
\citet{MarkoffFalckeFender2001} and \citet{Markoffetal2003}.  The
comparatively weak inverse Compton component in prior models followed
mainly from our choice of a small jet nozzle radius, $r_0\sim3 {\rm
r_{\rm g}}$, which was based on the assumption that the jet radius was on
the order of the event horizon.  If, on the other hand, the jet base
is contiguous with, or generated in, an extended corona, a larger
scale seems more sensible.  With this in mind, we here consider models
with with typical values of $r_0\gtrsim10r_{\rm g}$.

The dependence of the calculated spectrum upon the model parameters,
and their interdependence, are complex.  Increasing the scale of the
jet base decreases the electron density as well as the magnetic field,
for a fixed equipartition relationship.  This allows one to consider
electron temperatures several times higher than those used in our
previous models (to make up for lost synchrotron flux).  The higher
electron temperatures lead to greater inverse Compton emission
relative to synchrotron processes in the X-ray band.  Alternately, one
can compensate for the weakened radiating power of the decreased
particle density by increasing the total power input into the jet,
$Q_{\rm j}$.  Again, for a fixed equipartition relationship, the
particle density is more sensitive to this change than the magnetic
field (because the pressure of the field is $\propto B^2$) and thus
the inverse Compton component will experience a greater boost
relatively.


\begin{center}

\newcommand\tabspace{\noalign{\vspace*{0.7mm}}}
\newcommand\cyg{{Cyg~X-1}}
\def\errtwo#1#2#3{$#1^{+#2}_{-#3}$}

\begin{deluxetable}{rccccccccc} 
\setlength{\tabcolsep}{0.065in} 
\tabletypesize{\scriptsize} 
\tablewidth{0pt} 
\tablecaption{\gx\ and \cyg\ Jet Model Fits (90\% confidence level 
error bars) \label{tab:mffjetI}} 
\tablehead{ \colhead{Obs ID}  
           & \colhead{$N_{\rm j}$} 
           & \colhead{$^{\small a}L_{\rm j}$} 
           & \colhead{$r_{0}$}       
           & \colhead{$T_{\rm e}$}          
           & \colhead{$p$}              
           & \colhead{$k$}            
           & \colhead{$pl_{\rm f}$}  
           & \colhead{$z_{\rm acc}$} 
           & \colhead{$h_{0}$}      
           \\ 
           & ($10^{-3}$ L$_{\rm Edd}$)  
           & ($10^{-3}$ L$_{\rm Edd}$)  
           & ($GM/c^2$)  
           & ($10^{10}$\,K)  
          } 
\startdata 
20181-01-02 
            & \errtwo{1.27}{0.03}{0.02} 
            & \textsl{7.9}              
            & \errtwo{20.2}{0.3}{0.3}   
            & \errtwo{5.01}{0.07}{0.08} 
            & \errtwo{2.94}{0.01}{0.02} 
            & \errtwo{1.75}{0.09}{0.09} 
            & \errtwo{0.67}{0.03}{0.02} 
            & \textsl{25}               
            & \errtwo{1.34}{0.02}{0.02} 
            \\ 
\tabspace 
40108-02-01  
            & \errtwo{0.64}{0.02}{0.03} 
            & \textsl{4.7}              
            & \errtwo{9.6}{0.5}{0.1}    
            & \errtwo{5.23}{0.13}{0.12} 
            & \errtwo{2.39}{0.01}{0.00} 
            & \errtwo{1.12}{0.01}{0.01} 
            & \errtwo{0.74}{0.05}{0.01} 
            & \textsl{302}              
            & \errtwo{1.41}{0.01}{0.01} 
            \\ 
\tabspace 
40108-02-03  
            & \errtwo{0.34}{0.01}{0.03} 
            & \textsl{2.1}              
            & \errtwo{10.9}{0.8}{3.2}   
            & \textsl{4.0}              
            & \errtwo{2.65}{0.01}{0.00} 
            & \textsl{2.2}              
            & \textsl{0.7}              
            & \textsl{87}               
            & \errtwo{1.29}{0.03}{0.14} 
            \\ 
\tabspace 
\hline 
\tabspace 
40099-01-19  
            & \errtwo{0.85}{0.02}{0.03} 
            & \textsl{7.4}              
            & \errtwo{9.1}{0.2}{0.2}    
            & \errtwo{3.62}{0.05}{0.04} 
            & \errtwo{2.50}{0.03}{0.02} 
            & \errtwo{1.99}{0.18}{0.11} 
            & \errtwo{0.76}{0.02}{0.04} 
            & \textsl{16}               
            & \errtwo{1.33}{0.02}{0.02} 
            \\ 
\tabspace 
60090-01-26  
            & \errtwo{0.74}{0.01}{0.01} 
            & \textsl{7.1}              
            & \errtwo{4.4}{0.2}{0.1}    
            & \errtwo{3.28}{0.01}{0.01} 
            & \errtwo{2.61}{0.01}{0.01} 
            & \errtwo{1.77}{0.04}{0.01} 
            & \errtwo{0.73}{0.08}{0.02} 
            & \textsl{9}                
            & \errtwo{1.18}{0.00}{0.00} 
            \\ 
\tabspace 
60090-01-41  
            & \errtwo{0.78}{0.02}{0.04} 
            & \textsl{6.4}              
            & \errtwo{6.7}{0.1}{0.3}    
            & \errtwo{3.90}{0.03}{0.02} 
            & \errtwo{2.65}{0.00}{0.01} 
            & \errtwo{1.70}{0.01}{0.13} 
            & \errtwo{0.81}{0.03}{0.01} 
            & \textsl{14}               
            & \errtwo{1.17}{0.05}{0.00} 
            \\ 
\enddata 
 
\tablecomments{We failed to resolve 90\% confidence level error bars 
            for parameters listed in italics. Jet model parameters are 
            described in the paper and Appendix. Upper half of the 
            table refers to \gx, while the lower half refers to \cyg. 
            We fixed the mass, distance and inclination of \gx\ and 
            \cyg\ to 7 and 10 $M_\odot$, 6 and 2.5 kpc, and $30^\circ$ 
            and $47^\circ$, respectively 
            \citep{Hynesetal2003b,Hynesetal2004,Wuetal2001,Nowaketal1999}.\;$^a$Total output power from the jets in their rest frames $L_{\rm j}$ is 
            listed for completeness but is not itself a 
            fitted parameter (see discussion in the Appendix).} 
 
\end{deluxetable}

\begin{deluxetable}{rcccccccccl} 
\setlength{\tabcolsep}{0.065in} 
\tabletypesize{\scriptsize} 
\tablewidth{0pt} 
\tablecaption{\gx\ and \cyg\ Jet Model Fits, Continued \label{tab:mffjetII}} 
\tablehead{ \colhead{Obs ID}  
           & \colhead{$u_{\rm acc}/c$}  
           & \colhead{$f_{\rm sc}$} 
           & \colhead{L$_{\rm disk}$} 
           & \colhead{$T_{\rm disk}$}  
           & \colhead{$^{\small b}r_{\rm in}$} 
           & \colhead{$A_{\rm line}$} 
           & \colhead{$E_{\rm line}$}     
           & \colhead{$\sigma$}  
           & \colhead{$\Omega/2\pi$}  
           & \colhead{$\chi^2$/DoF}  
           \\ 
          &&& ($10^{-3}$ L$_{\rm Edd}$)  
           & (keV) 
           & ($r_{\rm g}$)  
           & ($10^{-2}$)  
           & (keV)           
           & (keV)  
         } 
\startdata 
20181-01-02  
            & \errtwo{0.79}{0.06}{0.09} 
            & \errtwo{1000}{70}{210}    
            & \errtwo{99}{1}{2}         
            & \errtwo{0.06}{0.16}{0.05} 
            & \textsl{486}              
            & \errtwo{0.30}{0.04}{0.03} 
            & \errtwo{6.3}{0.1}{0.1}    
            & \errtwo{1.0}{0.1}{0.1}    
            & \errtwo{0.14}{0.01}{0.03} 
            & 179/109  
            \\ 
\tabspace 
40108-02-01  
            & \errtwo{0.32}{0.05}{0.00} 
            & \errtwo{1100}{200}{800}   
            & \errtwo{0.33}{0.01}{0.01} 
            & \errtwo{1.53}{0.12}{0.10} 
            & \textsl{0.04}             
            & \errtwo{0.09}{0.03}{0.03} 
            & \errtwo{6.4}{0.1}{0.1}    
            & \errtwo{0.7}{0.2}{0.2}    
            & \errtwo{0.00}{0.06}{0.00} 
            &  118/87 
            \\ 
\tabspace 
40108-02-03  
            & \textsl{0.57}             
            & \textsl{230}              
            & \textsl{0.1}              
            & \textsl{0.36}             
            & \textsl{0.4}              
            & \errtwo{0.012}{0.006}{0.002} 
            & \errtwo{6.4}{0.2}{0.2}    
            & \errtwo{0.6}{0.2}{0.2}    
            & \errtwo{0.21}{0.11}{0.21} 
            &  22/36 
            \\ 
\tabspace 
\hline 
\tabspace 
40099-01-19  
            & \errtwo{0.55}{0.07}{0.01} 
            & \errtwo{940}{50}{200}     
            & \errtwo{2.4}{0.8}{0.6}    
            & \errtwo{0.71}{0.07}{0.06} 
            & \textsl{0.5}              
            & \errtwo{4.5}{0.4}{0.7}    
            & \errtwo{6.1}{0.2}{0.1}    
            & \errtwo{1.1}{0.1}{0.1}    
            & \errtwo{0.19}{0.02}{0.01} 
            &  225/154  
            \\ 
\tabspace 
60090-01-26  
            & \errtwo{0.35}{0.00}{0.00} 
            & \errtwo{790}{10}{10}        
            & \errtwo{0.8}{0.1}{0.1}    
            & \errtwo{0.98}{0.11}{0.09} 
            & \textsl{0.1}              
            & \errtwo{2.3}{0.2}{0.5}    
            & \errtwo{6.0}{0.1}{0.0}    
            & \errtwo{0.9}{0.1}{0.1}    
            & \errtwo{0.00}{0.01}{0.00} 
            &   207/177  
            \\ 
\tabspace 
60090-01-41  
            & \errtwo{0.38}{0.05}{0.01} 
            & \errtwo{710}{10}{170}     
            & \errtwo{0.8}{0.2}{0.4}    
            & \errtwo{0.88}{0.10}{0.03} 
            & \textsl{0.2}              
            & \errtwo{1.8}{0.3}{0.1}    
            & \errtwo{6.1}{0.1}{0.1}    
            & \errtwo{0.9}{0.2}{0.1}    
            & \errtwo{0.02}{0.01}{0.02} 
            &  186/179  
            \\ 
\enddata 
 
\tablecomments{We failed to resolve 90\% confidence level error bars 
            for parameters listed in italics. Jet model parameters are 
            described in the paper and Appendices.  Line parameters 
            correspond to the usual XSPEC/ISIS normalization, centroid 
            energy, and width of the {\tt gauss} line 
            model. $\Omega/2\pi$ is the reflection fraction from the 
            XSPEC/ISIS {\tt reflect} model.  Upper half of the table 
            refers to \gx, while the lower half refers to \cyg. 
            \;$^b$Values listed for inner disk radius $r_{\rm in}$ are 
            derived from the spectrally difficult-to-constrain fitted parameters  
            $L_{\rm disk}$ and $T_{\rm disk}$, 
            and are thus also unconstrained: see discussion in 
            Section~\ref{sec:results}.} 
 
\end{deluxetable}

\begin{deluxetable}{rccccccccccl}
\setlength{\tabcolsep}{0.065in}
\tabletypesize{\scriptsize}
\tablewidth{0pt}
\tablecaption{\gx\ and \cyg\ Coronal Model Fits (90\% Confidence Level
Error Bars) \label{tab:eqpair}}
\tablehead{ \colhead{Obs ID} 
           & \colhead{$A_{eqpair}$}    & \colhead{$\ell_{h}/\ell_{s}$} 
           & \colhead{$\tau_p$}        & \colhead{$A_{disk}$}
           & \colhead{$T_{disk}$}      & \colhead{$A_{line}$}
           & \colhead{$E_{line}$}      & \colhead{$\sigma$} 
           & \colhead{$\Omega/2\pi$}   & \colhead{$\xi$}
           & \colhead{$\chi^2$/DoF} 
           \\
           & ($10^{-3}$)               &
           &                           & ($10^{-5}$)
           & (keV)                     & ($10^{-2}$)
           & (keV)                     & (keV)
           &                           & ($4\pi F/n$) 
           & }
\startdata
20181-01-02 
            & \errtwo{0.21}{0.00}{0.07}   & \errtwo{5.85}{0.48}{0.33}
            & \errtwo{2.86}{0.25}{0.08}   & \errtwo{4.0}{1.4}{0.8}
            & \errtwo{1.30}{0.16}{0.24}   & \errtwo{0.14}{0.10}{0.04}
            & \errtwo{6.4}{0.1}{0.1}      & \errtwo{0.5}{0.4}{0.2}
            & \errtwo{0.11}{0.03}{0.04}   & \errtwo{680}{1460}{520}
            & 170/111
            \\
\tabspace
40108-02-01 
            & \errtwo{0.11}{0.03}{0.02}   & \errtwo{7.38}{0.55}{0.51}
            & \errtwo{1.79}{0.65}{0.59}   & \errtwo{1.1}{1.3}{0.8}
            & \errtwo{1.35}{0.16}{0.21}   & \errtwo{0.06}{0.04}{0.02}
            & \errtwo{6.4}{0.1}{0.1}      & \errtwo{0.4}{0.3}{0.4}
            & \errtwo{0.10}{0.04}{0.05}   & \errtwo{560}{3790}{560}
            & 117/90
            \\
\tabspace
40108-02-03 
            & \errtwo{0.03}{0.00}{0.09}   & \errtwo{3.73}{1.12}{0.28}
            & \errtwo{6.69}{0.92}{2.87}   & \errtwo{1.0}{0.8}{0.6}
            & \errtwo{1.10}{0.30}{0.31}   & \errtwo{0.02}{0.01}{0.01} 
            & \errtwo{6.4}{0.2}{0.2}      & \errtwo{0.7}{0.3}{0.4}
            & \errtwo{0.00}{0.26}{0.00}   & \nodata
            & 17/37
            \\
\tabspace
\hline
\tabspace
40099-01-19 
            & \errtwo{9.23}{0.01}{0.01}   & \errtwo{3.60}{0.01}{0.01}
            & \errtwo{1.35}{0.01}{0.01}   & \errtwo{422}{12}{3}
            & \errtwo{0.85}{0.01}{0.01}   & \errtwo{2.5}{0.1}{0.1}
            & \errtwo{6.2}{0.1}{0.0}      & \errtwo{0.8}{0.0}{0.1}
            & \errtwo{0.19}{0.01}{0.01}   & \errtwo{1140}{100}{170}
            & 226/158
            \\
\tabspace
60090-01-26 
            & \errtwo{6.78}{0.14}{2.11}   & \errtwo{5.88}{0.06}{0.07}
            & \errtwo{1.64}{0.02}{0.02}   & \errtwo{47}{12}{8}
            & \errtwo{0.87}{0.13}{0.00}   & \errtwo{1.9}{0.1}{0.2}
            & \errtwo{6.0}{0.1}{0.0}      & \errtwo{0.9}{0.0}{0.1}
            & \errtwo{0.22}{0.01}{0.01}   & \errtwo{2}{13}{2}
            & 191/181
            \\
\tabspace
60090-01-41 
            & \errtwo{6.09}{1.57}{1.19}   & \errtwo{5.67}{0.11}{0.05}
            & \errtwo{1.49}{0.02}{0.06}   & \errtwo{49}{13}{11}
            & \errtwo{0.85}{0.15}{0.04}   & \errtwo{1.4}{0.1}{0.2}
            & \errtwo{6.2}{0.0}{0.1}      & \errtwo{0.8}{0.1}{0.1}
            & \errtwo{0.22}{0.01}{0.01}   & \errtwo{2}{15}{2}
            & 186/183
            \\
\enddata

\tablecomments{Model parameters are for the XSPEC/ISIS implementations
            of the {\tt eqpair} model with ionized, smeared
            reflection, with additional contributions from a {\tt
            diskpn} and {\tt gauss} model. Upper half of the table
            refers to \gx, while the lower half refers to \cyg.}

\end{deluxetable}

\end{center}

\end{document}